\documentclass[doublecol]{epl2}
\usepackage{hyperref}
\usepackage{amsmath}
\usepackage{graphicx}

\title{Dynamical screening in strongly correlated metal SrVO$_{3}$}
\author{Li Huang\inst{1,2,3}\thanks{E-mail: \email{lihuang@iphy.ac.cn}} and Yilin Wang\inst{1,2}}
\shortauthor{Li Huang \etal}

\institute{                    
  \inst{1} Beijing National Laboratory for Condensed Matter Physics, 
           Chinese Academy of Sciences, 
           Beijing 100190, 
           China \\
  \inst{2} Institute of Physics, 
           Chinese Academy of Sciences, 
           Beijing 100190, 
           China \\
  \inst{3} Science and Technology on Surface Physics and Chemistry Laboratory, 
           P.O. Box 718-35, 
           Mianyang 621907, 
           China
}


\abstract{
The consequences of dynamical screening of Coulomb interaction among correlated 
electrons in realistic materials have not been widely considered before. In this 
letter we try to incorporate a frequency dependent Coulomb interaction into the 
state-of-the-art \emph{ab initio} electronic structure computing framework of 
local density approximation plus dynamical mean-field theory, and then choose 
SrVO$_{3}$ as a prototype material to demonstrate the importance of dynamical 
screening effect. It is shown to renormalise the spectral weight near the Fermi 
level, to increase the effective mass, and to suppress the $t_{2g}$ quasiparticle
band width apparently. The calculated results are in accordance with very recent 
angle-resolved photoemission spectroscopy experiments and Bose factor ansatz 
calculations.}

\begin{document}
\maketitle

\section{Introduction}

Over the last years, tremendous progresses have been made in the investigations 
of strongly correlated materials.~\cite{antoine:13,kotliar:865,va:2010} Such systems 
typically contain partially filled and relatively localized $3d$, $4f$ or $5f$ orbitals, 
and the low-energy physics of these systems is generally described by an effective 
Hamiltonian which models the electronic Coulomb interaction among these correlated 
orbitals. The electronic Coulomb interaction can be generally parameterized by a 
number, the ``Hubbard $U$". Usually $U$ is assumed to be static, but in principle 
$U$ is dynamical and frequency dependent: electric fields will be induced by the
charge density fluctuation in correlated orbitals, while the higher or lower-lying 
states in the systems will act to screen, resulting in a frequency dependent 
renormalization.~\cite{gw:1998,arya:125106,arya:195104} On one hand, it is realized 
that at high frequency the screening becomes weaker and eventually the interaction 
approaches the large bare Coulomb value, which is an order of magnitude larger than 
the static screened value. On the other hand, it is generally believed that the 
dynamical screening effect may play a key role in understanding the subtle electronic 
structures of strongly correlated materials,~\cite{casula:2012,werner:146401,casula:2011} for 
examples, plasmon satellites and spectral weight transfer in photoemission spectra 
of hole-doped BaFe$_{2}$As$_{2}$.~\cite{werner:331337} Thus how to take the dynamical 
screening effect into consideration is one of the challenges in studying the intriguing 
properties of strongly correlated materials.

In a realistic approach to strongly correlated materials, the frequency dependent 
$U = U(\omega)$ can be evaluated at the random phase approximation (RPA) level,
which allows for the direct evaluations of matrix elements of Hubbard $U$ and its
frequency dependence.~\cite{arya:125106,arya:195104,gw:1998} The static screened value 
$U_0 = \Re U(\omega = 0)$, has been widely used in the local density approximation plus 
static Coulomb interaction (LDA + $U$) method~\cite{ldau:943,amadon:155104} and 
local density approximation combined with dynamical mean-field theory (LDA + DMFT) 
approach.~\cite{va:2010,kotliar:865,antoine:13,amadon:205112,held:064202,
lechermann:125120} Since the frequency dependent screening has been completely 
neglected or empirically taken into account by adjusting the effective static $U_0$, 
very little is known on the consequences of frequency dependent Coulomb interaction 
in the electronic structures of strongly correlated materials. In order to include 
the dynamical screening effect into the LDA + DMFT framework, the hardest obstacle 
has been the lack of a reliable and efficient quantum impurity solver for the general 
impurity model with a frequency dependent Coulomb interaction $U(\omega)$.~\cite{werner:146401}

Fortunately, this obstacle has been recently overcome by the development of strong coupling 
continuous-time quantum Monte Carlo impurity solver (abbreviated CT-HYB) proposed by 
Werner \etal~\cite{werner:076405,werner:146401,werner:146404,gull:349}, where a multi-plasmon 
Lang-Firsov transformation is treated exactly in the context of a hybridization expansion 
algorithm for the general impurity model. By using this powerful impurity solver, Werner 
\etal\ have studied the plasmon satellites and large doping and temperature dependence of 
electronic properties in hole-doped BaFe$_{2}$As$_{2}$, a fascinating iron-based superconductor. 
They have demonstrated that the dynamical screening effect is important not only for 
high-energy spectral features, such as correlation satellites seen in photoemission 
spectroscopy, but also for the low-energy electronic structure, mass enhancements and 
lifetimes of hole-doped BaFe$_{2}$As$_{2}$.~\cite{werner:331337} As an very useful complement, 
the weak coupling continuous-time quantum Monte Carlo impurity solver (abbreviated CT-INT) 
developed by Rubtsov \etal~\cite{rubtsov:035122,gull:349,assaad:035116} can treat generic 
retarded interactions as well. But the CT-INT impurity solver is limited to small dynamical 
screened $U(\omega)$ and not-so-large screening frequencies $\omega$,~\cite{assaad:035116} 
and therefore becomes prohibitively costly for realistic applications of dynamical screening 
interactions in multi-orbital impurity models. M. Casula \etal~\cite{casula:2011} proposed 
a dubbed Bose factor ansatz (BFA) for the Green's function of quantum impurity problem with 
retarded interactions, in which the Green's function is factorized into a contribution 
stemming from an effective static-$U$ problem and a bosonic high-energy part introducing 
collective plasmon excitations. Various approximations for the Green's function Bose factor 
$F$ are introduced and their pros and cons are analyzed carefully. The most practical and 
effective one is borrowed directly from the so-called dynamical atomic limit approximation 
(DALA), whose functional form is analytically known. Since the DALA is valid in the 
hybridization $\Delta = 0$ limit, it only works well in the intermediate-strong coupling 
regime, with $U_{0}$ and the dynamical part large enough. The other functional forms going 
beyond the DALA scheme are available as well.~\cite{casula:2011}

Inspired by Werner \etal's pioneer works,~\cite{werner:331337,casula:2012} in this letter we 
try to incorporate the dynamical screening effect into the LDA + DMFT computing framework by 
adopting their approach.~\cite{werner:146401,werner:146404,werner:076405,gull:349} Then we 
apply this technique to solve a realistic multi-band Hamiltonian with dynamical Coulomb 
interaction $U(\omega)$ for prototypical transition metal oxide SrVO$_{3}$. The quasiparticle 
weight reduction, more renormalized $t_{2g}$ band width and the spectral weight transfer to 
higher frequencies are found and analyzed in details. Our calculations highlight the importance 
of including the proper screened interaction, to have a correct and reliable description of 
the electronic correlation in strongly correlated materials.

\section{Method}

The computational scheme used in present work can be viewed as an extension of the traditional 
LDA + DMFT method.~\cite{antoine:13,kotliar:865,va:2010} In a Hamiltonian formulation, the 
multi-orbital model with dynamically screened interactions can be written as
\begin{equation}
H = H^{\text{TB}}_{\text{LDA}} - H_{\text{DC}} + H_{\text{I}} + H_{\text{S}}.
\end{equation}
Here $H^{\text{TB}}_{\text{LDA}}$ is the single particle low-energy effective Hamiltonian. 
To generate $H^{\text{TB}}_{\text{LDA}}$, the ground state calculation by using the 
pseudopotential plane-wave method with the {\scriptsize ABINIT} package~\cite{abinit:2582} 
has been performed at first. For the ground state calculation, the projector augmented 
wave (PAW) type pseudopotentials~\cite{paw:17953} for Sr, V and O species 
are constructed by ourselves, the cutoff energy for plane-wave expansion is 20\ Ha, 
and the $k$-mesh for Brillouin zone integration is $ 12 \times 12 \times 12 $. These 
pseudopotentials and computational parameters are carefully checked and tuned to ensure 
the numerical convergences. And then $H^{\text{TB}}_{\text{LDA}}$ is obtained by applying 
a projection onto maximally localized Wannier function (MLWF)~\cite{wann:12847} 
orbitals including all the V-$t_{2g}$ and O-$2p$ orbitals, which resulting in a $12 \times 12$ 
tight-binding Hamiltonian. $H_{\text{DC}}$ is the double counting term, and the around 
mean-field  (AMF) scheme~\cite{amadon:155104} is used, which is especially suitable for 
strongly correlated metal system. The Coulomb interaction term $H_{\text{I}}$ is taken 
into consideration merely among three V-$t_{2g}$ orbitals. In the present work, we choose 
$U$ = 4.0 and $J$ = 0.65\ eV, which are close to previous 
estimations.~\cite{amadon:205112,lechermann:125120} The bosonic part $H_{\text{S}}$ reads 
\begin{equation}
H_{\text{S}} = \sum_{i} \int \text{d}\omega 
\left[\lambda_{i\omega} (b^{\dagger}_{i\omega} + b_{i\omega})\sum_{\alpha}n_{i\alpha} 
+ \omega b^{\dagger}_{i\omega}b_{i\omega} \right],
\end{equation}
which describes the coupling of electronic degrees of freedom (occupation number $n$) to 
bosonic modes (bosonic operators $b^{\dagger}$ and $b$) and the coupling strength 
$\lambda_{\omega} = \sqrt{\Im U(\omega)/\pi}$. The bosonic modes represent 
the screening of the charge density fluctuations on the strongly correlated 
V-$t_{2g}$ orbitals.

The multi-orbital lattice problem is studied using DMFT,~\cite{antoine:13,kotliar:865} 
which maps it to a self-consistent solution of a three-orbital quantum impurity model. 
Then the impurity model (with frequency-dependent interactions $U(\omega)$) is solved using the 
CT-HYB impurity solver~\cite{werner:146401,werner:146404,werner:076405,gull:349} which 
based on a stochastic diagrammatic expansion of the partition function in the 
impurity-bath hybridization. In fact, the effect of $U(\omega)$ is to dress the fermionic 
propagators with a bosonic propagator $\exp{[-\mathcal{K}(\tau)]}$ where $\mathcal{K(\tau)}$ 
is the twice-integrated retarded interaction. In terms of $\Im U(\omega)$
and a factor $\mathcal{B}(\tau,\omega) = \cosh[(\tau-\frac{\beta}{2})\omega] / \sinh[\frac{\beta\omega}{2}]$ 
with bosonic symmetry, $\mathcal{K}(\tau)$ can be expressed by~\cite{werner:146401,werner:331337}
\begin{equation}
\label{eq:k}
\mathcal{K}(\tau) = \int^{\infty}_{0} 
\frac{d\omega}{\pi} 
\frac{\Im U(\omega)}{\omega^2}
[\mathcal{B}(\tau,\omega)-\mathcal{B}(0,\omega)].
\end{equation}
Thus the diagrammatic weight for Monte Carlo stochastic sampling is supplemented with 
an additional bosonic term, which originates from the contribution of $\mathcal{K}(\tau)$. 
The implemented details of CT-HYB impurity solver improved with recently developed 
orthogonal polynomial representation algorithm~\cite{boe:075145} can be easily found in the
literatures.~\cite{werner:146404,werner:146401,werner:076405} In each LDA + DMFT iterations, 
typically $4 \times 10^{8}$ Monte Carlo samplings have been performed to reach sufficient 
numerical accuracy. All the calculations are carried out at the inverse temperature of 
$\beta = 10$ or $\beta = 30$. Finally the maximum entropy method~\cite{jarrell:133} is 
used to perform analytical continuation to obtain the impurity spectral functions of 
V-$t_{2g}$ states.

\section{Results and discussion}

SrVO$_{3}$ is a prototype of correlated metal and very well studied material which
represents a benchmark for theories describing strongly correlated compounds.~\cite{casula:2011}
Indeed, its band structure is relatively simple due to its undistorted perovskite 
structure, resulting in the occupation of one electron in three-fold degenerate 
V-$t_{2g}$ bands crossing the Fermi level. The Oxygen $2p$ bands are quite well 
separated from the $t_{2g}$ levels, such that the $3 \times 3$ V-$t_{2g}$ tight-binding 
Hamiltonian is a minimal model required for a correct description of the low-energy 
physics of SrVO$_{3}$. Thus, SrVO$_{3}$ has been the testing case for many new LDA + DMFT 
implementations.~\cite{amadon:205112,lechermann:125120,held:064202,
nekrasov:155112,pavar:176403,nek:155106,lee:165103} And on the other hand, SrVO$_{3}$ 
has been the subject of intensive experimental activity, with magnetic, electrical, 
optical measurements, and by means of angle-resolved photoemission spectroscopy 
(ARPES).~\cite{yos:146404,yos:085119,eguchi:076402,sekiyama:156402} Therefore SrVO$_{3}$ 
is an ideal system to benchmark our newly developed implementation of LDA + DMFT 
computing framework incorporated with dynamical screening interaction.

\begin{figure}
\centering
\includegraphics[scale=0.35]{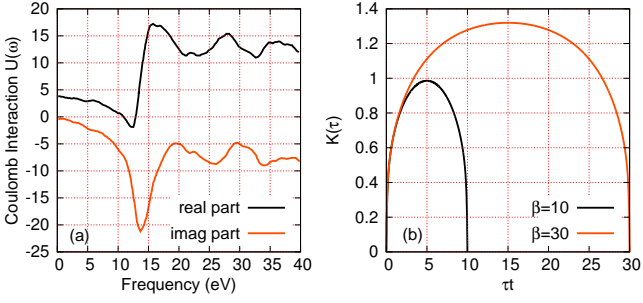}
\caption{(Color online) Frequency dependent interaction for SrVO$_{3}$
from constrained random phase approximation. (a) Real and imaginary
parts of the average intra-orbital Coulomb interaction as a function of 
frequency. The data shown in this figure are taken from reference~\cite{arya:125106} 
directly. (b) The twice-integrated retarded interaction function $\mathcal{K}(\tau)$
at finite temperatures $\beta=10$ and $\beta=30$ calculated by 
Eq.(\ref{eq:k}).\label{fig:scr}}
\end{figure}

To include dynamical screening interaction within LDA + DMFT computing scheme, one of 
the bottlenecks is the determination of the frequency dependent Coulomb interaction 
$U(\omega)$ and corresponding twice-integrated retarded interaction $\mathcal{K}(\tau)$.
Recent works try to extract these quantities from constrained density functional 
calculations,~\cite{ldau:943} GW-inspired methods,~\cite{gw:1998} or RPA-based 
schemes.~\cite{arya:125106,arya:195104} All these calculations are very tedious and 
tricky. Fortunately, the realistic frequency dependent $U(\omega)$ for SrVO$_{3}$ has 
been computed before~\cite{arya:125106} based on the constrained RPA (cRPA) 
approach, so we can extract $U(\omega)$ directly from the reference as a necessary 
input to evaluate $\mathcal{K}(\tau)$ in the preprocessing stage. The obtained 
$U(\omega)$ and $\mathcal{K}(\tau)$ are shown in Fig.\ref{fig:scr} respectively. Clearly 
seen in Fig.\ref{fig:scr}(a), $\Im U(\omega)$ features prominent peaks at 14\ eV, 26\ eV 
and 33\ eV respectively. From the peaks of $\Im U(\omega)/\omega^{2}$, the energies of 
plasmon satellite structures in the V-$t_{2g}$ density of states can be easily concluded.
The real part of $U(\omega)$ ranges from the static screened value $U_{0} = 4.0$\ eV 
to the bare unscreened value of about 12\ eV at large $\omega$, and the frequency 
dependence resembles typical $U(\omega)$ in strongly correlated metals. In a standard 
LDA + DMFT calculation without dynamical screening effect, the relatively small value 
of static screened value $U_{0}$ would result in a rather weakly correlated picture.

\begin{figure}
\centering
\includegraphics[scale=0.65]{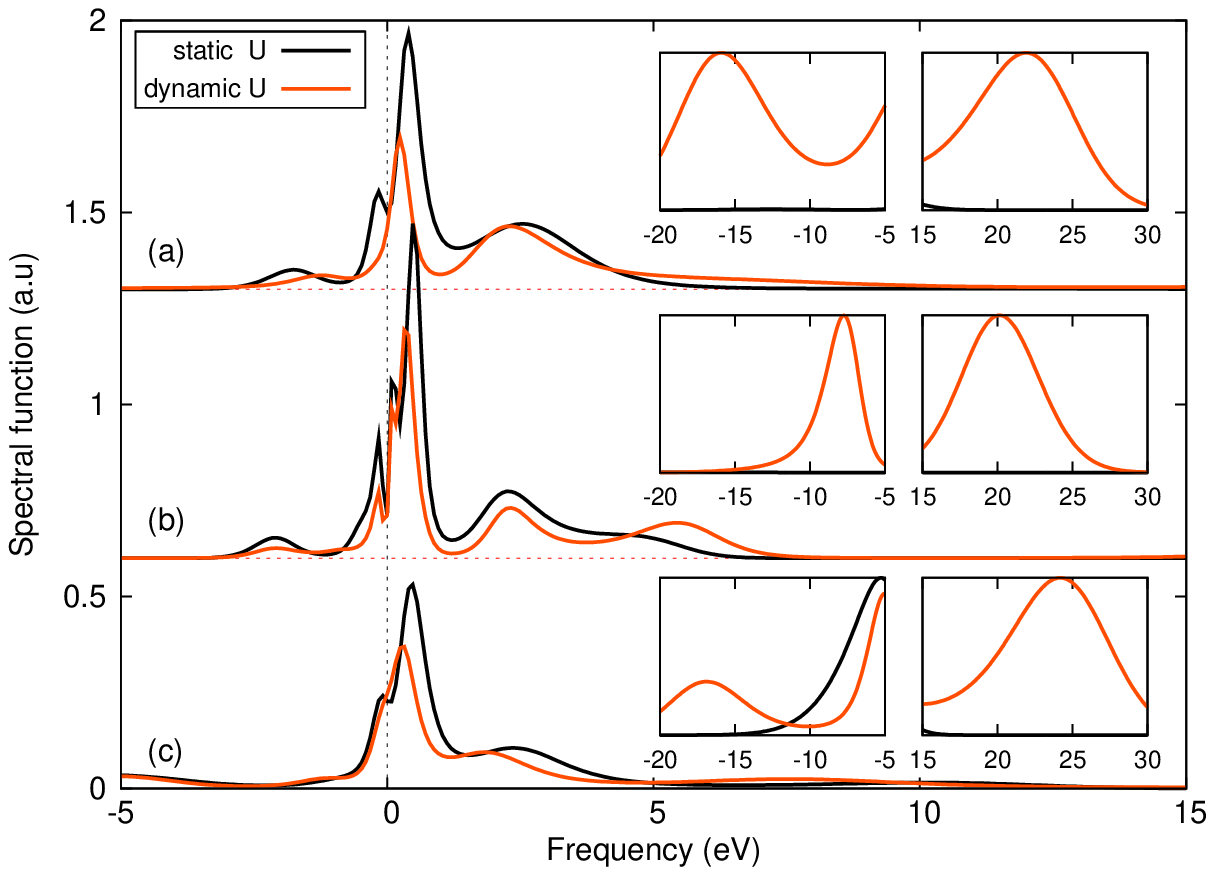}
\includegraphics[scale=0.65]{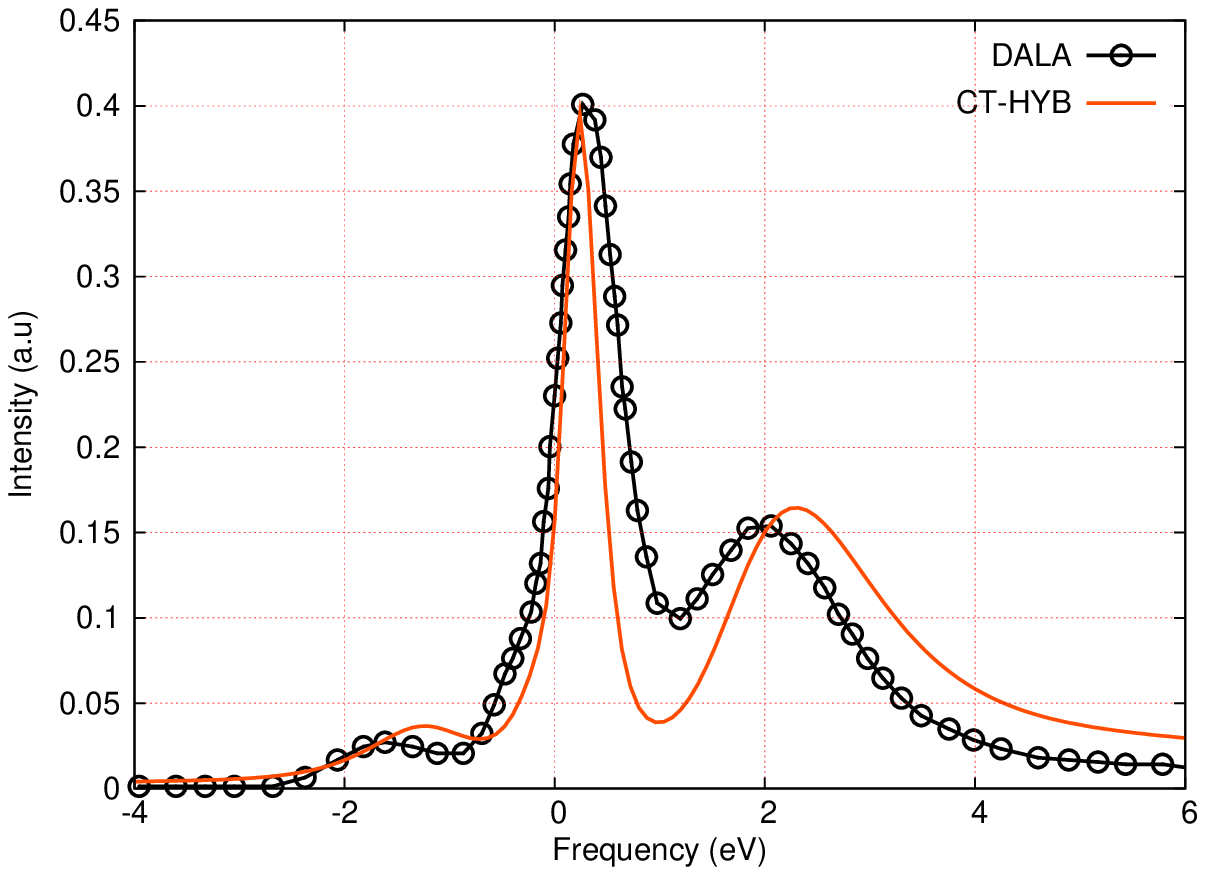}
\caption{(Color online) Integrated spectral functions $A(\omega)$ of V-$t_{2g}$ states. 
Upper panel: Spectral functions obtained in current LDA + DMFT calculations for different 
models. (a) $N_{\text{wann}}=3$, $\beta=10$; (b) $N_{\text{wann}}=3$, $\beta=30$; 
(c) $N_{\text{wann}}=12$, $\beta=10$. Inset: The fine structures of spectral functions 
at high frequencies. The spectral functions are obtained from imaginary-time Green's 
functions $G(\tau)$ by using maximum entropy method,~\cite{jarrell:133} and the results 
are cross-checked by using recently developed stochastic analytical continuation 
method.~\cite{beach} Lower panel: Comparison of spectral functions of V-$t_{2g}$ states 
obtained by our LDA + DMFT calculations with CT-HYB impurity solver and BFA-DALA method 
respectively. $N_{\text{wann}} = 3$, $\beta = 10$. The results obtained by BFA-DALA method 
are extracted directly from the reference~\cite{casula:2011}.\label{fig:spectrum}}
\end{figure}

Now let's focus on the spectral properties of V-$t_{2g}$ states. Two different models for SrVO$_{3}$
are taken into considerations, one contains only the V-$t_{2g}$ states (number of Wannier orbitals 
$N_{\text{wann}} = 3$), the other contains both V-$t_{2g}$ and O-$2p$ states (number of Wannier orbitals 
$N_{\text{wann}} = 12$). We solve these models in the framework of LDA + DMFT at $\beta = 10$ and $\beta = 30$
with dynamical $U$ and static $U$ respectively. The calculated results are illustrated in the upper panel 
of Fig.\ref{fig:spectrum}. The results obtained by dynamical $U$ and static $U$ calculations display 
remarkable differences. Firstly, in the dynamical $U$ calculations, spectral weights are shifted 
strongly to high frequencies, which leads to a reduction in weight at low energies, compared with 
the static $U$ results. Secondly, the quasiparticle resonance peak around the Fermi level is strongly 
renormalized, not only the width but also the height of it shrink apparently as $U(\omega)$ is taken 
into account. Thirdly, in dynamical $U$ calculations, some shoulder peaks near $\omega = 0$ are 
smeared out. Thus, the explicit treatment of the Coulomb repulsion at large frequencies has a substantial 
effect, even on the low-energy properties of the system.~\cite{casula:2011,werner:146401,werner:331337}

We note that recently Casula \etal~\cite{casula:2011} have studied the spectral properties of SrVO$_{3}$
by using their BFA-DALA method. Since both the BFA-DALA method and our CT-HYB scheme are intent to deal 
with the strongly correlated systems with frequency dependent Coulomb interaction, it should be meaningful 
to justify the calculated results obtained by these two different methods. Comparison of the calculated 
spectral functions of V-$t_{2g}$ states are shown in the lower panel of Fig.\ref{fig:spectrum}. Both 
spectral functions exhibit significant three-peak structures, which are prominent for strongly correlated 
metal systems.~\cite{antoine:13} Clearly, the spectral function obtained by our CT-HYB scheme is consistent 
with that obtained by BFA-DALA method, besides a slightly spectral weight reduction near the Fermi level 
and a small shift of the upper Hubbard band to high energies. These discrepancies are certainly due to the 
smaller value of static screened $U$ ($U_{0} = 3.6$\ eV) used in Casula \etal's BFA-DALA calculations.\cite{casula:2011}

\begin{figure}
\centering
\includegraphics[scale=0.35]{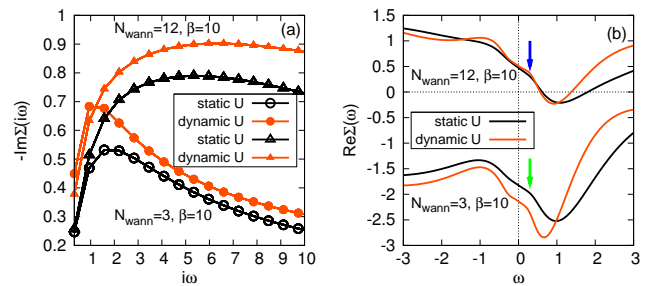}
\caption{(Color online) Self-energy function of V-$t_{2d}$ states obtained by LDA + DMFT
calculations at finite temperature. Left panel:Imaginary part of self-energy function at matsubara 
frequency axis $\Im\Sigma(i\omega)$. Right panel:Real part of self-energy function at real axis 
$\Re\Sigma(\omega)$. The blue and green arrows are used to indicated the ``kink"-like 
structures, which are often considered as signatures of strongly correlated systems.\label{fig:sigma}}
\end{figure}

In Figure \ref{fig:sigma} the calculated self-energy function of V-$t_{2g}$ states are reported.
Let's concentrate our attention to the imaginary part of self-energy function at matsubara frequency 
axis at first. In dynamical $U$ model, $-\Im \Sigma(i\omega \rightarrow 0)$ is remarkable larger 
than that of corresponding static $U$ model. According to the well-known Eliashberg equation:~\cite{antoine:13} 
$Z^{-1} \approx 1 - \frac{\beta}{\pi}\Im\Sigma(i\omega \rightarrow 0)$, the quasiparticle weight $Z$ 
can be approximately evaluated. Thus the dynamical screening effect results in a smaller value 
of $Z$, namely a larger value of effective mass $m^{*}$. Indeed, for the realistic dynamical $U$
model we found a $Z \approx 0.43 \sim 0.48$, which gives an effective mass renormalized by $2.1 
\sim 2.3$ with respect to the LDA band structure. On the other hand, for the corresponding 
static $U$ model we obtained a value of $Z \approx 0.6$, which underestimates the electronic 
correlation by a factor of 1.67, and so the value of $m^{*}$. Recent ARPES data yielded an 
effective mass $m^{*} \approx 2m_{0}$,~\cite{yos:146404,eguchi:076402,yos:085119} which is 
in good agreement with our findings for the realistic dynamical interaction. It is worthy 
noting that a static $U$ model with a larger instantaneous $U_{0}$~\cite{nekrasov:155112,
pavar:176403,nek:155106} can be used to fit the experimental mass renormalization artificially. 
But the lower Hubbard band peaked at $-1.5$\ eV can not be obtained by such a static model. The 
difficulty here is to reproduce both the effective mass and the position of the lower Hubbard 
band by the same model. With the dynamically screened interaction model, we can describe 
correctly both the effective mass ($\sim 2.1m_{0}$) and the lower Hubbard band peaked at 
$\sim -1.5$ eV, as is seen in both Fig.\ref{fig:sigma} and Fig.\ref{fig:spectrum}. 

After careful analytical continuation, self-energy function of V-$t_{2g}$ states at real axis 
is obtained and plotted in Fig.\ref{fig:sigma}(b). For the real part of self-energy function 
$\Re\Sigma(\omega)$, there are two 
extrema at the energies $\omega \sim \pm 1.0$ eV, originating from the crossover from the central
quasiparticle resonance peak to the lower and upper Hubbard bands. In the energy regime of the 
quasiparticle resonance peak ranging from about -0.5\ eV to 0.5\ eV, $\Re\Sigma(\omega)$ can be
roughly described by a straight line and the slope of it can be used to evaluate 
the quasiparticle weight $Z$ as well. Not surprisingly, the quasiparticle weight Z calculated 
with $\Re\Sigma(\omega)$ is almost identical with that calculated with $\Im\Sigma(i\omega)$.
In this energy range the Landau Fermi liquid theory is valid.
Strictly speaking, the Fermi liquid regime with $\Re\Sigma(\omega) \sim -\omega$ only extends from 
-0.2 up to 0.2\ eV, which is in accord with previous LDA + DMFT calculations.~\cite{nekrasov:155112} 
Next to this Fermi liquid regime, there are pronounced ``kink"-like structures in $\Re\Sigma(\omega)$ 
around $\omega = \pm 0.25$\ eV as are highlighted by colourful arrows. Apparently, the ``kink"-like 
structures are enhanced by the dynamical screening effects. These ``kink"-like structures can 
be regarded as a fingerprint of strongly correlated systems~\cite{by:168} and become important in 
the context of quasiparticle dispersion and electronic specific heat. In previous LDA + DMFT 
calculation~\cite{nekrasov:155112} and ARPES experiment~\cite{yos:146404} the ``kink"-like 
structures in SrVO$_{3}$ have been already confirmed.

\begin{figure}
\centering
\includegraphics[scale=0.65]{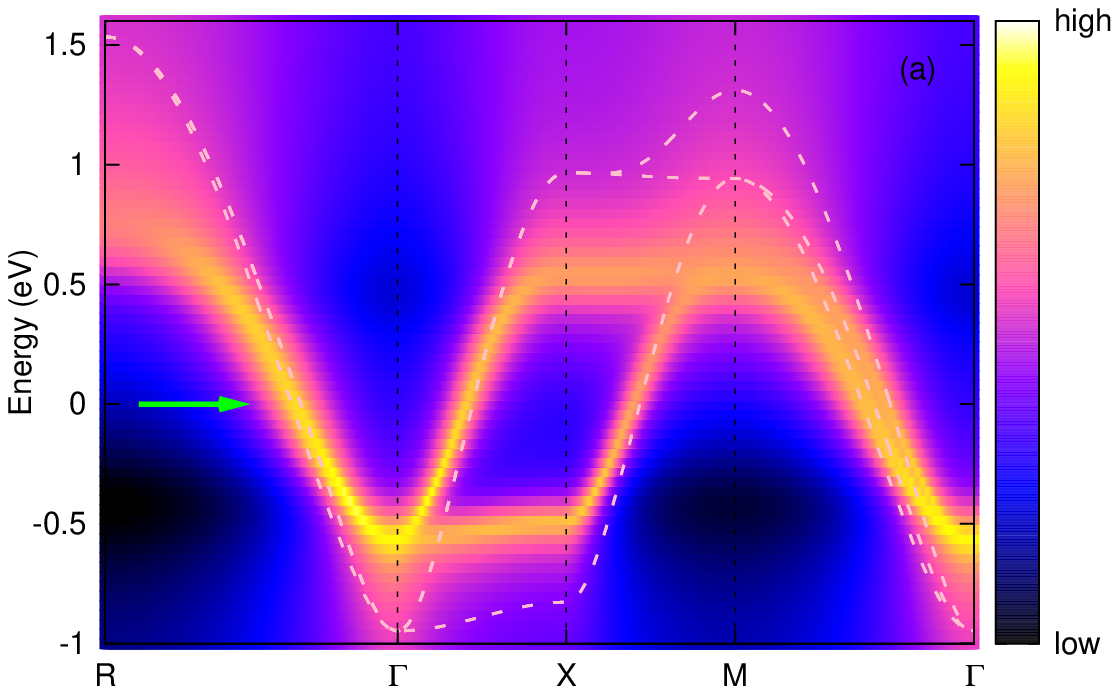}
\includegraphics[scale=0.65]{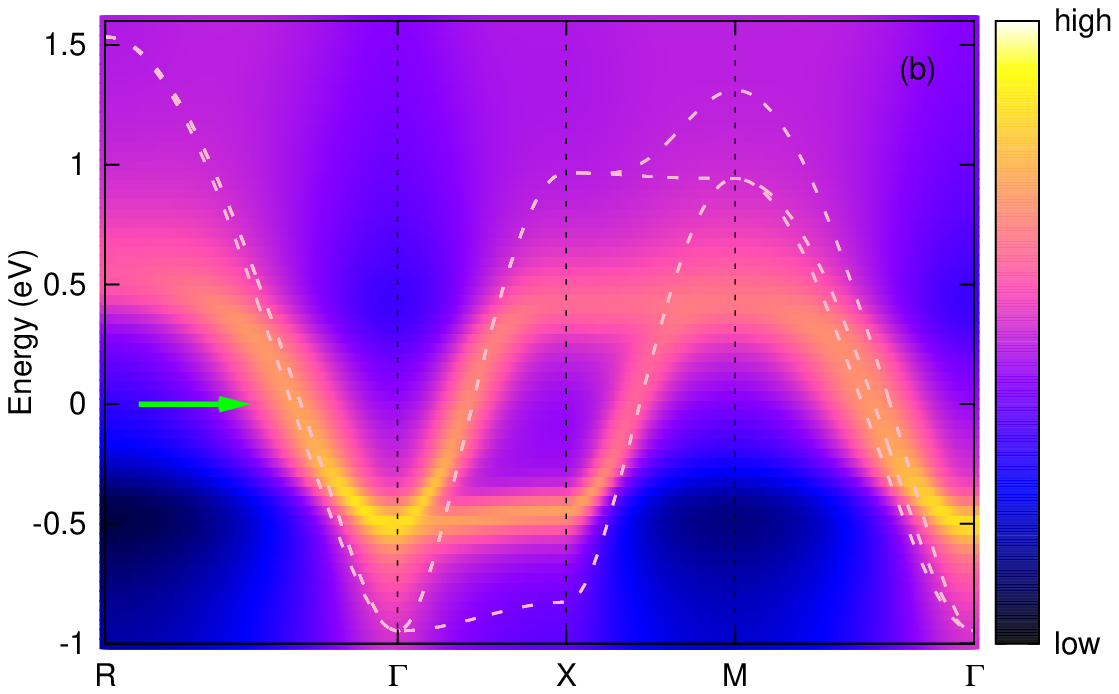}
\caption{(Color online) Angle-resolved spectral functions $A(\text{k},\omega)$ for
SrVO$_{3}$. Upper panel: Static Coulomb interaction case. Lower panel: Frequency dependent
Coulomb interaction case. The colourful dash lines denote LDA band structures and the 
green arrows indicate ``kink"-like structures. In these LDA + DMFT calculations, only
three V-$t_{2g}$ bands are taken into considerations and the inverse temperature $\beta$ 
is fixed to 10.\label{fig:arpes}}
\end{figure}

With the knowledge of self-energy function on the real axis, we are now in
the position to calculate k-resolved spectral functions or quasiparticle 
dispersion $A(\text{k},\omega)$. The LDA + DMFT quasiparticle dispersions 
for dynamical $U$ and static $U$ models along high symmetry lines in the 
Brillouin zone are shown in Fig.\ref{fig:arpes} respectively. The LDA band 
structure is presented in this figure as an useful comparison. As for 
the static $U$ model, the quasiparticle band structure displays strongly 
band renormalization with a factor $\sim 1.7$ with respect to LDA band 
dispersion. For example, we can see from the upper panel of Fig.\ref{fig:arpes} 
that the bottom of the quasiparticle band is located at approximately 
$\omega = -0.6$\ eV, in contrast to the LDA value of $\omega = -1.0$\ eV. 
Near the Fermi level along $R - \Gamma$ and $M - \Gamma$ lines there exist 
discernible ``kink"-like structures as indicated by a green 
arrow,\cite{nekrasov:155112,yos:146404} which 
stem from the shoulders in the real part of the self-energy function 
(see Fig.\ref{fig:sigma}(b)). As for the dynamical $U$ model, its 
quasiparticle band structure exhibits similar features with respect to that 
of static $U$ model. Around the Fermi level, the quasiparticle band structure 
of dynamical $U$ model is more renormalized with a factor $\sim 2.1$ and 
has less intensity, and the bottom of quasiparticle dispersion is located at 
$\omega = -0.5$\ eV. The ``kink" structures in quasiparticle band 
structure are more discernible than those of corresponding static $U$ model.

\section{Conclusion}
In this letter, we try to incorporate the dynamical screening effect into the modern
LDA + DMFT computing framework with CT-HYB as a quantum impurity solver. We apply
this calculation scheme to reinvestigate the electronic structure of SrVO$_{3}$ with
dynamical $U$ model. It seems that the dynamical screening effect will enhance 
the electron correlation significantly. The transfer of spectral weight to higher 
frequencies, further reduction of quasiparticle resonance peak, smaller quaisparticle weight 
$Z$, larger effective mass $m^{*}$, more apparent ``kink"-like structures etc. are 
predicted or verified by our calculations as well. Based on the calculated results, 
it is suggested that the dynamical screening interaction may play an important role 
in understanding the fine electronic structures of strongly correlated materials, 
which has been ignored by most of previous theoretical calculations.

\acknowledgments
We acknowledge financial support from the National Science Foundation ͑of China and that
from the 973 program of China under Contract No.2007CB925000 and No.2011CBA00108. All the 
LDA + DMFT calculations have been performed on the SHENTENG7000 at Supercomputing Center of 
Chinese Academy of Sciences (SCCAS).

\bibliographystyle{eplbib}
\bibliography{screen}

\end{document}